\documentclass[manuscript]{aastex631}

\begin{document}

\title{FAST detection of OH emission in the carbon-rich planetary nebula  NGC\,7027}

\correspondingauthor{Yong Zhang}
\email{zhangyong5@mail.sysu.edu.cn}

\author[0000-0002-2762-6519]{Xu-Jia Ouyang}
\affiliation{School of Physics and Astronomy, Sun Yat-sen University, 2 Daxue Road, Tangjia, Zhuhai, Guangdong Province,  People's Republic China}

\author[0000-0002-1086-7922]{Yong Zhang}
\affiliation{School of Physics and Astronomy, Sun Yat-sen University, 2 Daxue Road, Tangjia, Zhuhai, Guangdong Province,  People's Republic China}
\affiliation{Xinjiang Astronomical Observatory, Chinese Academy of Sciences, 150 Science 1-Street, Urumqi, Xinjiang 830011, People's Republic of China}
\affiliation{CSST Science Center for the Guangdong-Hongkong-Macau Greater Bay Area, Sun Yat-Sen University, Guangdong Province, People's Republic China}

\author[0000-0002-4428-3183]{Chuan-Peng Zhang}
\affiliation{National Astronomical Observatories, Chinese Academy of Sciences, Beijing 100101, People's Republic China}
\affiliation{Guizhou Radio Astronomical Observatory, Guizhou University, Guiyang 550000, People's Republic China}

\author{Peng Jiang}
\affiliation{National Astronomical Observatories, Chinese Academy of Sciences, Beijing 100101, People's Republic China}
\affiliation{Guizhou Radio Astronomical Observatory, Guizhou University, Guiyang 550000, People's Republic China}

\author[0000-0003-3324-9462]{Jun-ichi Nakashima}
\affiliation{School of Physics and Astronomy, Sun Yat-sen University, 2 Daxue Road, Tangjia, Zhuhai, Guangdong Province,  People's Republic China}
\affiliation{CSST Science Center for the Guangdong-Hongkong-Macau Greater Bay Area, Sun Yat-Sen University, Guangdong Province, People's Republic China}
                
\author[0000-0002-5435-925X]{Xi Chen}
\affiliation{Center for Astrophysics, Guangzhou University, Guangzhou 510006, People's Republic China}
\affiliation{Shanghai Astronomical Observatory, Chinese Academy of Sciences, 80 Nandan Road, Shanghai 200030,
People's Republic China}

\author[0000-0003-0196-4701]{Hai-Hua Qiao}
\affiliation{National Time Service Center, Chinese Academy of Sciences, Xi'An, Shaanxi 710600, People's Republic of China}
\affiliation{Key Laboratory of Time Reference and Applications, Chinese Academy of Sciences, People's Republic China}

\author{Xu-Ying Zhang}
\affiliation{School of Physics and Astronomy, Sun Yat-sen University, 2 Daxue Road, Tangjia, Zhuhai, Guangdong Province,  People's Republic China}

\author[0009-0007-4663-2643]{Hao-Min Sun}
\affiliation{School of Physics and Astronomy, Sun Yat-sen University, 2 Daxue Road, Tangjia, Zhuhai, Guangdong Province,  People's Republic China}

\author[0000-0003-2090-5416]{Xiao-Hu Li}
\affiliation{Xinjiang Astronomical Observatory, Chinese Academy of Sciences, 150 Science 1-Street, Urumqi, Xinjiang 830011, People's Republic of China}

\author[0000-0002-3171-5469]{Albert Zijlstra}
\affiliation{Department of Physics and Astronomy, The University of Manchester, Manchester M13 9PL, UK}



\begin{abstract}

{We present the first detection of the ground-state OH emission line at 1612\,MHz toward the prototypical carbon-rich planetary nebula (PN) NGC\,7027, utilizing the newly installed ultra-wideband (UWB) receiver of the Five-hundred-meter Aperture Spherical radio Telescope (FAST). This emission is likely to originate from the interface of the neutral shell and the ionized region.} The other three ground-state OH lines at 1665, 1667, and 1721\,MHz are observed in absorption and have velocities well matched with that of HCO$^+$ absorption. We infer that the OH absorption is from the outer shell of NGC\,7027, although the possibility that they are associated with a foreground cloud cannot be completely ruled out. All the OH lines exhibit a single blue-shifted component with respect to the central star. The formation of OH in carbon-rich environments  might be via photodissociation-induced chemical processes. Our observations offer significant  constraints for chemical simulations, and they underscore the potent capability of  the UWB receiver of FAST to search for nascent PNe.

\end{abstract}

\keywords{Astrophysical masers (103); Hydroxyl masers (771); Planetary nebulae (1249); Circumstellar masers (240)}

\section{Introduction} \label{sec:intro}

 Planetary nebulae (PNe), originating from low- and intermediate-mass stars, represent significant objects for {material exchange between
stars and the interstellar medium.}
 In the asymptotic giant branch (AGB) phase, stars become unstable and expel material into space at a mass-loss rate up to 10$^{-5}$--10$^{-4}$\,$M_\sun$\,yr$^{-1}$ \citep{1993ApJ...413..641V,1995A&A...299..755B,2019NatAs...3..408D}.
{Following a brief post-AGB phase \citep[$10^2 - 10^4$ yr;][]{2016A&A...588A..25M},} the effective temperature (T$_{\rm eff}$) of the central star increases rapidly to $\sim$30\,kK, initiating the ionization of the surrounding circumstellar envelope (CSE) by its radiation, thereby instigating the formation of PN \citep[][]{1993ARA&A..31...63K}.
 Evolved stars are classified as carbon-rich (C/O $>$ 1), S-type (C/O $\approx$ 1), or oxygen-rich (C/O $<$ 1) based on the carbon to oxygen ratio in their CSEs \citep{2018A&ARv..26....1H}.

Maser emissions  of SiO, H$_2$O, and OH molecules are frequently detected in the oxygen-rich AGB CSEs \citep{1996A&ARv...7...97H}. 
However, as the AGB evolves into the PN phase, these masers gradually become extinct.
It is anticipated that maser emissions, such as SiO, H$_2$O, and OH, will sequentially disappear, typically dying out $\sim$10, $\sim$100, and $\sim$1000 years after the end of AGB mass-loss. 
Consequently, the presence or absence of different maser species can serve as temporal indicators, effectively delineating distinct stages of stellar evolution \citep{1989ApJ...338..234L,1990RMxAA..20...55G}.
The ground-state OH transitions originate from four hyperfine sublevels within the lowest rotational state $^2\prod_{3/2}$, $J=3/2$, exhibiting frequencies of 1612.231 ($F=1\rightarrow2$), 1665.402 ($F=1\rightarrow1$), 1667.359 ($F=2\rightarrow2$), and 1720.530\,MHz ($F=2\rightarrow1$), respectively.
Through the interaction of excitation and de-excitation of rotational excited states ($^2\prod_{3/2}, J=5/2$ and $^2\prod_{1/2}, J=1/2$), the relative populations of hyperfine sublevels can deviate from their intrinsic values in local thermodynamic equilibrium (LTE)
\citep[e.g.,][]{2000A&A...353.1065H,2015ApJ...815...13E}.
The OH maser emission at these frequencies exhibit distinct characteristics. 
Specifically, the 1612\,MHz OH maser is predominantly detected within the CSE of evolved stars \citep[e.g.,][]{1979A&A....75..351N,2012A&A...547A..40U}, whereas the OH masers from the main-line transitions at 1665 and 1667\,MHz are mainly associated with massive star-forming regions \citep{2000ApJS..129..159A,2018ApJS..239...15Q}. 
The maser  at 1721\,MHz has primarily been observed in supernova remnants \citep[SNRs; e.g.,][]{1996AJ....111.1651F}.
While the 1612\,MHz OH maser has been commonly detected in evolved stars, it gradually annihilates with the evolution of the central star.
As a result, compared to AGB CSEs, PNe rarely exhibit OH masers.
 Thus far, only eight PNe have been detected or possibly detected 
in OH maser emission \citep{1989A&A...217..157Z,1991A&A...243L...9Z,2012A&A...547A..40U,2016ApJ...817...37Q},
which are sometimes called OHPNe or nascent PNe\footnote{ {The eight OHPNe 
are Vy\,2–2, IRAS\,17393–2727, K\,3–35, IRAS\,17347–3139, JaSt\,23, IRAS\,16333–4807, NGC\,6302, and IRAS\,07027–7934, 
among which the }  1612\,MHz OH line detected in  NGC\,6302 may be dominated by thermal emission 
\citep{2016MNRAS.461.3259G,2020ApJS..247....5Q}, casting some doubt on
its OHPN nature.}.
Except IRAS\,07027--7934 \citep{1991A&A...243L...9Z} all these PNe are oxygen-rich.
Recently, four additional sources exhibiting OH maser emission were classified as OHPN candidates by matching the interferometric position of OH maser emission with the radio continuum emission position observed in the Southern Parkes Large-Area Survey in Hydroxyl  \citep[SPLASH,][]{2022MNRAS.516.2235C}. Investigating the behavior of OH lines is pivotal
for understanding the early evolution of PNe.

In this paper, we report the first detection of ground-state OH 
lines toward the carbon-rich PN NGC\,7027.
NGC\,7027 is a prototypical young PN and one of the brightest and well-studied object \citep[e.g.,][]{1967ApJ...149L..97G,1973ApL....15...87B,1980ApJ...238..892M}.
Situated at a distance of $\sim$890 pc \citep{2015Ap&SS.357...21A}, it possesses a dynamical age of $\sim$1000 years \citep{2008ApJ...681.1296Z,2018A&A...609A.126S}.
Photometric studies reveal a hot central star with an effective temperature of $\sim200$\,kK \citep{2000ApJ...539..783L,2005A&A...442..249Z,2023ApJ...942...15M}, exhibiting a ultraviolet (UV) luminosity of $\sim3 \times 10^{37}$ erg\,s$^{-1}$ \citep{2000ApJ...539..783L,2023ApJ...942...15M} and an X-ray luminosity of $\sim7 \times 10^{31}$ erg\,s$^{-1}$ \citep{2012AJ....144...58K,2018ApJ...861...45M}.
Optical and infrared observations show that NGC\,7027 is a carbon-rich nebula \citep[C/O=2.7,][]{2005A&A...442..249Z}, implying that its progenitor has a mass of $\sim2$--4\,M$_\sun$ \citep[][and references therein]{2021ApJ...922...24K}.
{The infrared spectroscopy reveals
prominent bands from aromatic hydrocarbons,
supporting its carbon-rich nature
\citep{1996A&A...315L.369B,2001A&A...367..949B}.}
Observations of its kinematics revealed an asymmetrical expansion, suggesting the presence of multiple  outflows within the nebula \citep{2000ApJ...539..783L,2002A&A...384..603C,2012RMxAA..48....3L,2016ApJ...833..115L}.
The systematic velocity of NGC\,7027 is $\sim$24.3\,$\rm km~s^{-1}$ 
 \citep{2023ApJ...942...14B}. It was quite unexpected when 
 \cite{1996A&A...315L.257L} discovered the  far-infrared rotational 
 lines of H$_2$O and OH in the Infrared Space Observatory spectrum of
this carbon-rich PN. The presence of H$_2$O was subsequently confirmed 
by the observation with Herschel Space Observatory
\citep{2010A&A...518L.144W}.

\section{Observations and Data Reduction} \label{sec:obse}

The observations were carried out on November 18th 2022, 
employing the newly installed cryogenic ultra-wideband (UWB) receiver of the Five-hundred-meter Aperture Spherical radio Telescope \citep[FAST;][]{2022RAA....22k5016L,2023RAA....23g5016Z}. 
The ON-OFF position switching mode was used with a total integration time of 1 hour. The pointing position was set at $\rm R.A. = 21^h07^m01^s.57$, $\rm decl. = +42\arcdeg14\arcmin10\arcsec.5$ (J2000), with the OFF position $\sim$20$\arcmin$ away from the ON position.
The UWB receiver operates within the frequency range of 500$-$3300\,MHz and is segmented into four subbands, each comprising 1,048,576 channels, providing a frequency resolution of 1049.04\,Hz.
The velocity resolution achieved for the ground-state OH lines is $\sim$0.2\,$\rm km~s^{-1}$. 
We smoothed the spectra to a resolution of $\sim$1.5\,$\rm km~s^{-1}$ to improve the signal-to-noise ratio by averaging over more data points.
The half-power beam width (HPBW) is $\sim 2.5 \arcmin$ at 1650\,MHz.
Periodically, a high-intensity noise of $\sim$12\,K is injected for flux calibration.

{After examining the time-frequency diagram, we eliminated the radio frequency interference, and} obtained the antenna temperature ($T_{\rm A}$) through
\begin{equation}
T_{\rm A} = T_{\rm cal}\frac{P^{\rm cal}_{\rm on}}{P^{\rm cal}_{\rm on}-P^{\rm cal}_{\rm off}},
\end{equation}
where $T_{\rm cal}$ is the injected temperature of the noise diode, 
and $P^{\rm cal}_{\rm on}$ and $P^{\rm cal}_{\rm off}$ denote the power values when the diode is on and off, respectively.
The flux density $S_{\nu}$ can be deduced through $S_{\nu} = T_{\rm A}/G$, where $G$ is the gain \citep[see,][for details]{2023RAA....23g5016Z}.
The linear polarizations XX and YY were calibrated and subtracted with
the baseline separately, and then were combined
to obtain the final spectra.
The detection was deemed genuine only if its flux density exceeded the 3$\sigma$ root-mean-square (rms) noise level, the signal extended across more than multiple adjacent channels, and both linear polarizations XX and YY displayed the signal.

\section{Results} \label{sec:resu}

\begin{figure}[ht!]
\plotone{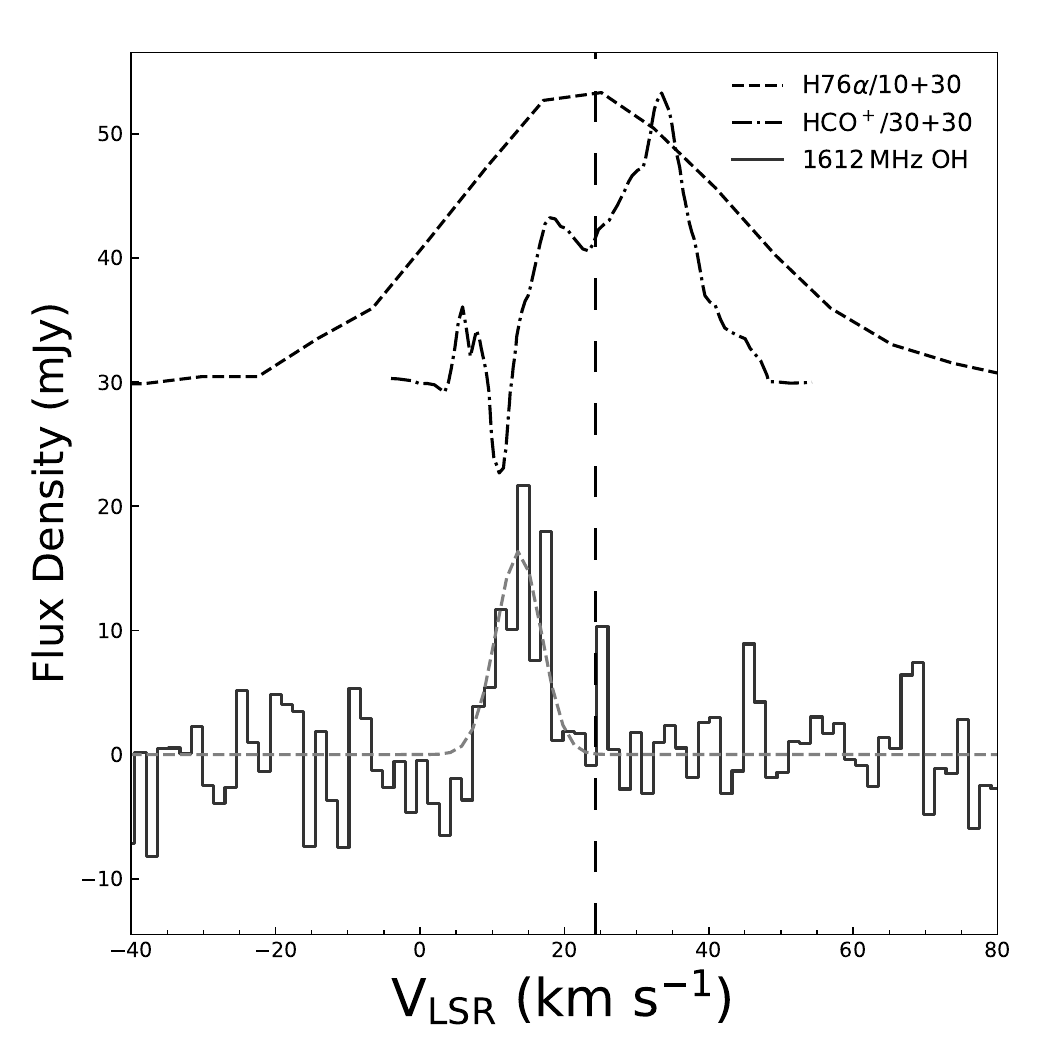}
\caption{
The OH emission line at 1612\,MHz (solid binned line). 
V$_{\rm LSR}$ is the velocity in the Local Standard of Rest.
The gray dashed curve represents the Gaussian fitting result. The vertical straight line indicates  the systemic velocity. For comparison, we overplot the profiles of the HCO$^+$ and
H76$\alpha$ lines taken from 
\citet{2023ApJ...942...14B} and \citet{1991A&A...251..611R}, respectively.
\label{fig:emi}
}
\end{figure}

\begin{figure}[ht!]
\plotone{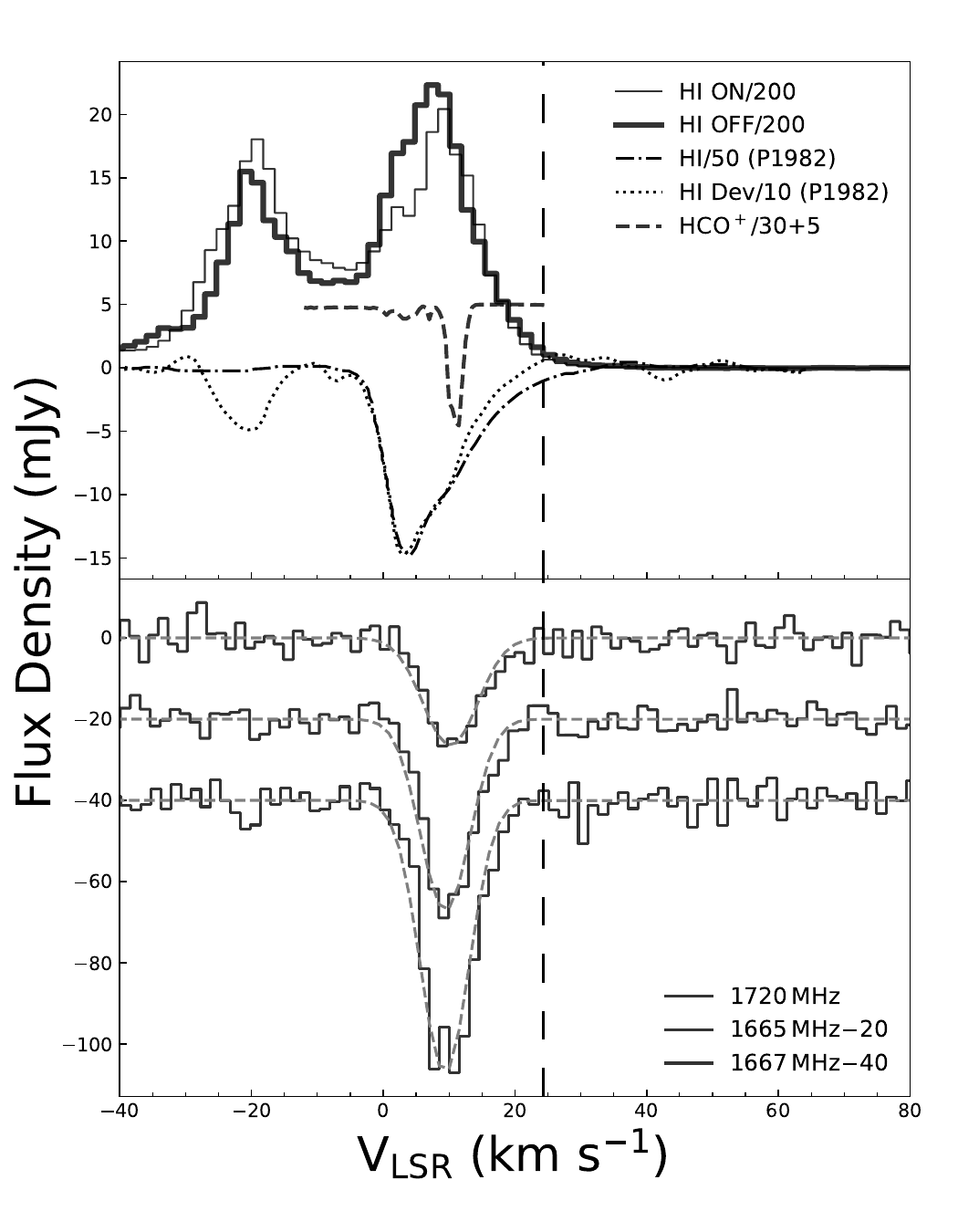}
\caption{Lower panel: Same as Figure~\ref{fig:emi} but
for the OH absorption lines at 1665, 1668, and 1720\,MHz. Upper panel: The ON- and OFF-postion
spectra of \ion{H}{1}. For comparison, we
overplot the HCO$^+$ (dashed curves) and \ion{H}{1} absorption spectra taken from \citet{2023ApJ...942...14B} and 
\cite{1982A&A...106..229P},
where the 
dot-dashed and dotted curves represent 
toward NGC\,7027 and an extragalactic source, 
respectively. 
\label{fig:abs}}
\end{figure}

As shown in Figure~\ref{fig:emi} and Figure~\ref{fig:abs}, we clearly detect an OH emission line at 1612\,MHz and OH absorption lines at 1665, 1667, and 1721\,MHz toward the
ON position. The behavior of these OH features is strikingly similar to
that found in NGC\,6302 \citep{1988ApJ...326..368P}, and is opposite
to that in most of Galactic OH sources where the 1612\,MHz line is
observed in absorption while the 1721\,MHz line is in emission \citep{2020MNRAS.497.4066H,2022MNRAS.512.3345D}.
These features do not appear in the OFF-position spectrum.
The 1612\,MHz OH emission in NGC\,7027 is fainter than those reported for known OHPNe. All four OH lines show a single component. 
Gaussian functions were utilized to fit the OH features for the
measurements. The fitting results are outlined in Table \ref{tab:pro},
where the integrated fluxes, the  width at half maximum (FWHM),
the peak velocities (V$_{\rm P}$) and fluxes (S$_{\rm P}$),
and the noise root-mean-square (rms) values are given.
The central velocities of the main lines
and the 1612\,MHz emission line
suggest an expanding velocity of 14.9\,km~s$^{-1}$ and 10.6\,km~s$^{-1}$,
respectively, indicating that the absorption and emission may come from
different regions.

NGC\,7027 has been included in the sample of the OH survey performed
by \cite{1988ApJ...326..368P} utilizing the NRAO 43-meter telescope,
but no firm detection was claimed. As the NRAO 43-meter telescope
has  a HPBW of $\sim$18$\arcmin$ at 1650\,MHz, the beam-diluted intensity
is about 50 times weaker than that of the FAST spectrum. Therefore,
\cite{1988ApJ...326..368P} were unlikely to see any OH signal in their
lower sensitivity spectrum.

It is imperative to validate that the OH emission and absorption indeed originate from this PN rather than from interstellar gas.
We scrutinize the databases of masers presented by 
\citet{2015A&A...582A..68E} and \citet{2019AJ....158..233L},
and do not find any record within a radius of $\sim17\arcmin$ surrounding NGC\,7027. Are the OH features associated with a foreground
\ion{H}{1} cloud? The wide frequency coverage of UWB allows to observe
the \ion{H}{1} 21\,cm line simultaneously. We detect two \ion{H}{1}
emission lines peaking around $-20$ and 10\,km~s$^{-1}$ in both
ON- and OFF-position spectra, as shown in Figure~\ref{fig:abs}, which
are from two extensive clouds.
The interferometer observations of \citet{1982A&A...106..229P} show
that the two \ion{H}{1} lines appear to be absorption against an
extragalactic source and the dip at $-20$\,km~s$^{-1}$ disappeared
when the telescope points toward NGC\,7027 (see Figure~\ref{fig:abs}), suggesting that the
positive- and negative-velocity clouds are in front of and beyond NGC\,7027, respectively. 
It is arguable whether the OH absorption is associated with the foreground \ion{H}{1} cloud.
If so, it is challenging to comprehend that \ion{H}{1} is seen in emission while OH is seen 
in absorption since FAST has similar beam sizes at 18 and 21\,cm, though it is not impossible. Given that its velocity is different than the cloud's, the OH emission at 1612\,MHz is most likely to originate from the circumstellar shell.

If these OH lines originate from the circumstellar shell, their peak
velocities suggest an expansion velocity of 10--15\,km~s$^{-1}$,
which is consistent with that observed through the nebular CO line profile \citep[see the references in][]{2000ApJ...532..994H}.
The inner ionized gas has a higher expansion velocity, as indicated by
the H76$\alpha$ line (Figure~\ref{fig:emi}; note that
it also has a thermal broadening of $\sim15$\,km~s$^{-1}$).
The interaction between the fast ionized gas and the surrounding
neutral materials generates a dense shell, where the
photodissociation region (PDR)  is situated.
\cite{2023ApJ...942...14B} discover a slower HCO$^+$ emission shell and
a HCO$^+$ absorption at $\sim10$\,km~s$^{-1}$, which are associated with the PN.
They infer that HCO$^+$ originates outside the PDR. 
Figure~\ref{fig:emi} shows that the OH emission fits very well with the inner edge of the HCO$^+$ shell, and thus is likely to be from the PDR.
Figure~\ref{fig:abs} shows the absorption
component of HCO$^+$ that is obtained
by summing the negative pixels over a 
area $13\arcsec\times13\arcsec$ in the image
of \cite{2023ApJ...942...14B}.
It appears that the HCO$^+$ and  OH absorption
come from the same velocity component,
probably suggesting that the OH absorption lines originate from the cold layer outside the PDR.
The \ion{H}{1} absorption spectrum of \citet{1982A&A...106..229P} may display
excess absorption at $\sim10$\,km~s$^{-1}$  compared to the extragalactic source, which
could be tentatively attributed to a circumstellar component. High angular resolution interferometer observations are needed to validate these findings.
In addition, the HCO$^+$ spectrum has a shallow dip at $\sim3$\,km~s$^{-1}$, which should be
from the foreground \ion{H}{1} cloud.

\section{Discussion} 

\begin{figure}[ht!]
\plotone{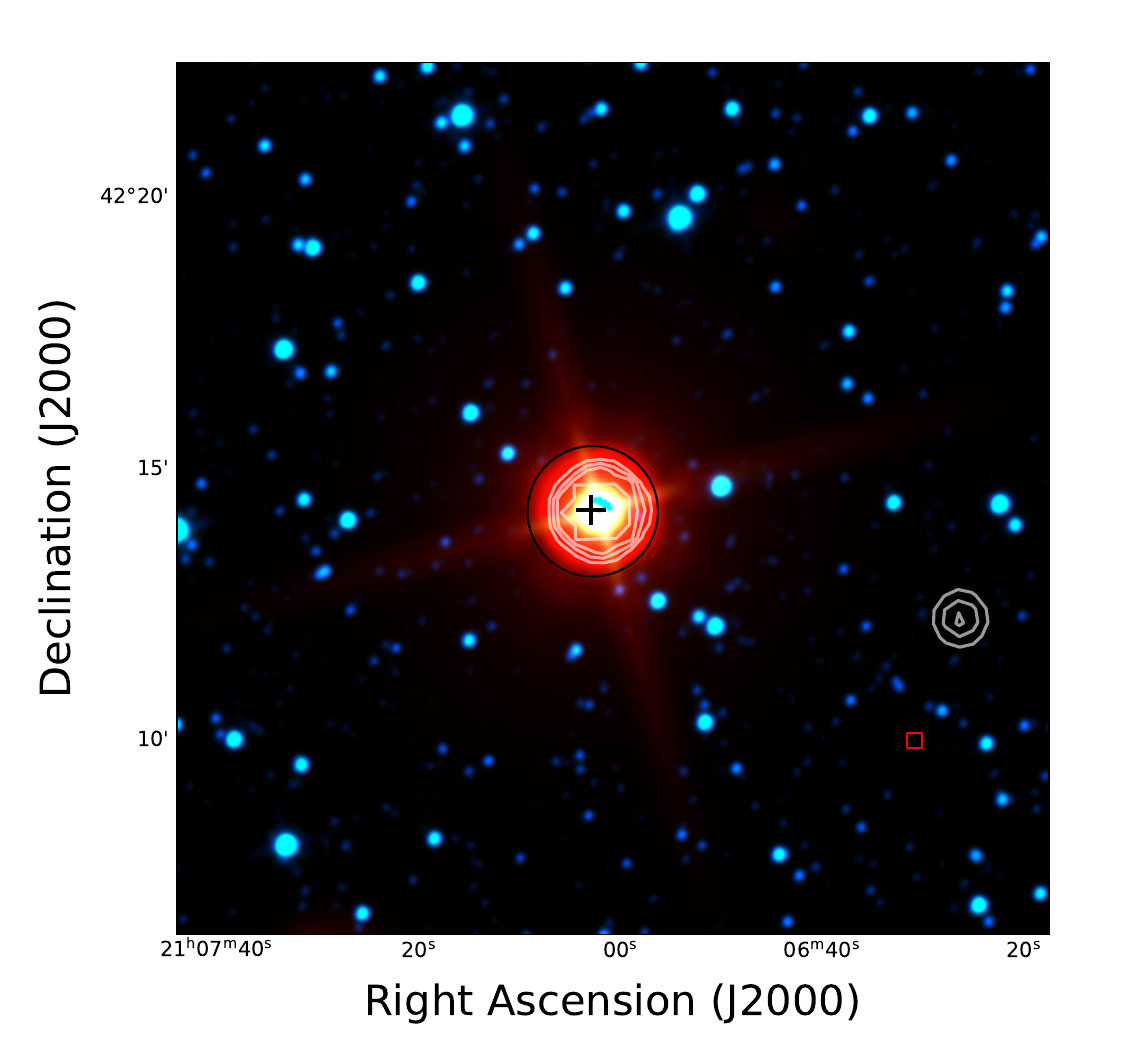}
\caption{The WISE image of NGC\,7027. The 3.4, 4.6, and 22\,$\mu m$ bands are 
shown in blue, green, and red, respectively. 
The black circle and cross indicate the beam size of FAST at 1650\,MHz and the 
beam center.
The red square denotes the
\ion{H}{2} region in this field.
The contour of the 1.4\,GHz continuum emission
is overplotted.
\label{fig:wise}}
\end{figure}

OH masers toward AGB stars  typically exhibit a double-peaked profile,
where the blue- and red-shifted components  respectively originate from the approaching and receding sides of the CSE \citep{1997A&AS..122...79S}.
{In contrast,} OH emission in OHPNe displays a pronounced asymmetry, with a notable preference for blue-shifted emission respect to the systemic velocity of the central star \citep{1989A&A...217..157Z,2012A&A...547A..40U,2016ApJ...817...37Q}.
The blue-shifted OH emission can be attributed to either the
obscuration of the red-shifted emission by the ionized gas or the 
amplification of the continuum emission by the foreground gas. The absence of the red-shifted
OH emission accords with the optically thick
nature of NGC\,7027 at 18\,cm. 
Figure~\ref{fig:wise} shows the infrared images surrounding  NGC\,7027 obtained with the Wide-field Infrared Survey Explorer (WISE) space telescope,
in which the continuum emission from the NRAO VLA Sky Survey \citep[NVSS,][]{1998AJ....115.1693C} is overlaid. NGC\,7027 is the strongest infrared
and radio emission source within the field,
facilitating the production of blue-shifted OH lines.

The 1612 and 1721\,MHz OH satellite lines may have a conjugate character with one observed in emission while the other observed
in absorption \citep{1976ApJ...203..124E,1978A&A....66..395G}.
Such conjugate lines are commonly detected in the sources with bright continuum emission background \citep[eg.,][]{2005Sci...309..106W,2016MNRAS.461.3259G,2022MNRAS.512.3345D}.
The decay of the second rotationally excited level 
can enhance the population of the
$F=1$ levels on the ground state, resulting in an emission at
1612\,MHz  and an absorption at 1721\,MHz.
The pumping is probably done by far-infrared radiation
\citep{2019ApJ...871...89E}. An alternative mechanism is chemical
pumping in the hot PDR where OH is formed and destroyed rapidly,
in which and the inversion comes from the decay from the high excitation
levels after formation \citep{1968Natur.217..334S}. Comprehensive theoretical analysis of 
the excitation of these OH lines in PN circumstances is sorely lacking.

Under an assumption of LTE, the beam-average column density of OH molecules can be estimated from  the main absorption lines using the equation given
by \citet{1996A&A...314..917L},
\begin{equation}
N_{\rm OH} = C_0 \times \Delta T_{\rm ex} \int \frac{T_A}{\eta} dv  \, \rm cm^{-2},
\end{equation}
where $C_0 = 4 \times 10^{14}$ and $2.24 \times 10^{14}$\,$\rm cm^{-2} K^{-1} km^{-1} s$ for the 1665 and 1667 MHz transitions respectively \citep{1968ApJS...15..131G},
and the main beam efficiency $\eta$ is 0.8--0.9 \citep{2021RAA....21..282S}.
$\Delta T_{\rm ex}$ is defined by
\begin{equation}
\Delta T_{\rm ex} = \frac{T_{\rm ex}}{T_{\rm ex}-T_{\rm bg}},
\end{equation}
where $T_{\rm ex}$ and $T_{\rm bg}$
is the excitation temperature of the OH
main lines and the background continuum temperature, respectively.
For an order of magnitude estimate we assume $T_{\rm ex}$ = 50\,K
and $T_{\rm bg}$ = 1000\,K. The resultant $N_{\rm OH}$ is about
$10^{14}$\,cm$^{-2}$. Given a H$_2$ column density of 
$N_{\rm H2}=(7.7 \pm 3.3) \times 10^{21}$\,cm$^{-2}$ \citep{2008ApJ...678..328Z},
we obtain a $N_{\rm OH}/N_{\rm H2}$ abundance ratio of
$\sim10^{-8}$, which is comparable to the typical value in
diffuse clouds \citep{2012A&A...542L...7W}. 

NGC\,7027 provides only the second case of OH  emission in a carbon-rich PN after IRAS\,07027--7934 \citep{1991A&A...243L...9Z}.\footnote{The excited-state OH maser at 4765\,MHz has been detected in a carbon-rich CSE CRL\,618
\citep{2019ApJ...878...90S}, which, however, is usually regarded as a proto-PN.} The presence of oxygen-bearing molecules in carbon-rich environment has two
potential explanations. Oxygen-bearing molecules may be located in a fossil oxygen-rich envelope that has been ejected before the PN progenitor was transformed into a carbon star. 
{If that is the case, OH should be mostly distributed in the outer shell.
According to the CO map of NGC\,7027 \citep{2023ApJ...942...14B},
the molecular shell has a diameter of $\sim30\arcsec$.
The mass of H$_2$ in of NGC\,7027 is $\sim 1$\,M$_\sun$ \citep{2012A&A...545A.114S}. Assuming a OH/H$_2$ abundance ratio
of $10^{-8}$, the column density of OH is roughly estimated to be 
$\sim10^{11}$\,cm$^{-2}$, about three orders of magnitude lower
than the observed value.
Therefore, the extended AGB remain envelope 
is unlikely to produce strong OH absorption as observed here. }

Instead, the detection of OH
infrared lines suggest that this molecule is freshly synthesized in
the hot PDR of NGC\,7027 through
an endothermic reaction $\rm O + H_2 \rightarrow OH + H$
\citep{1996A&A...315L.257L}, where O is released from CO  
photodissociation. This scenario is supported by chemical models
of NGC\,7027 \citep{1999ApJ...515..640Y,2000ApJ...532..994H}.
The column density and fractional abundance of OH derived form our observation are in reasonably good agreement with the modelling predictions.
OH could also be enhanced through the photodissociation of 
H$_2$O,  which is evidently present in NGC\,7027 \citep{2010A&A...518L.144W}.
However, the formation of H$_2$O in situ in the PDR requires an extremely high temperature ($\sim3500$\,K), and the formed OH can be 
rapidly destroyed at these temperatures, either reacting back to H$_2$O or dissociating in a reaction with H.
Recently, \citet{2022MNRAS.510.1204V} show that
H$_2$O may be present already in the wind of a carbon-rich AGB star
if there is UV radiation from a binary companion 
\footnote{{The AGB envelope contains a substantial amount of dust so that interstellar UV photons cannot penetrate it sufficiently to trigger the photochemistry
\citep[see e.g.,][]{1997A&A...324..237W} }.}. 
This could provide
a plausible explanation for the presence of circumstellar OH.
Due to its extreme  obscuration, it is unclear whether 
the central star of NGC\,7027 has a binary companion.
{To account for the rings and jets detected in NGC\,7027, 
\citet{2023ApJ...942...15M} suggest the presence of an unseen
binary campanion. The detection of circumstellar OH might provide 
a further evidence for that hypothesis.
}

\section{Conclusion} \label{sec:concl}

Using the UWB receiver of FAST, we  detect the 1612\,MHz emission line and 1665, 1667, 1721\,MHz absorption lines of OH toward the carbon-rich PN NGC\,7027.
We have compelling evidence supporting the circumstellar origin of the
OH emission although we cannot completely rule out the possibility that
the the OH absorption lines may be associated with a foreground cloud.
These OH ground-state lines exhibit a single blue-shifted component,
and behave similarly to those  in NGC\,6302 albeit with much lower
intensities. The detection of OH is uncommon among PNe, and could
shed important insights into the transition from the post-AGB to PN
phase. Elaborated excitation models are needed to understand the physical 
conditions of the OH region, for which our observations provide 
stringent constrains. 
 A search for OH in a larger sample utilizing FAST is expected to substantially increase the number of  OHPNe.

\section*{Acknowledgements}

{We acknowledge an anonymous referee for his/her constructive comments.}
We are very grateful to Jesse Bublitz who helpfully provided us with the the absorption spectrum of HCO$^+$. The financial supports of this work are from the Guangdong Basic and Applied Basic Research Funding (No.\,2024A1515010798),
the National Natural Science Foundation of China (NSFC, No.\,12333005), and the science research grants from the China Manned Space Project (NO. CMS-CSST-2021-A09, CMS-CSST-2021-A10, etc).
Y.Z. thank the Xinjiang Tianchi Talent Program (2023).
H.-H.Q. is supported by the Youth Innovation Promotion Association CAS and the NSFC(No.\,11903038). A.A.Z acknowledges support from STFC/UKRI through grant ST/X001229/1. 
FAST is operated and managed by the National Astronomical Observatories, Chinese Academy of Sciences.


\bibliography{sample631}{}
\bibliographystyle{aasjournal}



\newpage

\clearpage
\newpage

\begin{deluxetable*}{ccccccc}[htb]
\tablewidth{0pt}
\linespread{0.8}
\tablecaption{Measurements of the OH emission and absorption toward NGC\,7027. \label{tab:pro}}
\tablehead{\\
\colhead{Frequency} & \colhead{$\int S_v dv$} & \colhead{$V_{\rm p}$} & \colhead{FWHM} & \colhead{$S_{\rm p}$} & \colhead{rms}\\
\colhead{(MHz)} & \colhead{($\rm mJy~km~s^{-1}$)} & \colhead{($\rm km~s^{-1}$)} & \colhead{($\rm km~s^{-1}$)} & \colhead{(mJy)} & \colhead{(mJy)} & \\
\colhead{(1)} & \colhead{(2)} & \colhead{(3)} & \colhead{(4)} & \colhead{(5)} & \colhead{(6)}
}
\startdata
1612 & 128.2$\pm$21.8  & 13.7$\pm$0.6 & 7.4$\pm$1.4 & 16.4$\pm$2.8 & 3.3 \\
1665 & $-$455.6$\pm$16.4  & 9.4$\pm$0.2 & 8.8$\pm$0.4 & $-$48.4$\pm$1.7 & 2.7 \\
1667 & $-$638.9$\pm$25.0  & 9.4$\pm$0.2 & 8.7$\pm$0.4 & $-$68.7$\pm$2.7 & 3.1 \\
1721 & $-$278.0$\pm$15.2  & 10.2$\pm$0.3 & 9.7$\pm$0.6 & $-$26.8$\pm$1.5 & 2.8 \\
\enddata
\end{deluxetable*}

\end{document}